\documentclass[aps,prb,twocolumn,superscriptaddress,showpacs,floatfix]{revtex4}
\usepackage{graphicx}
\usepackage{bm}

\begin{document}

\title{``Unusual'' critical states in type-II superconductors}

\author{E.~H.~Brandt}
\affiliation{Max-Planck-Institut f\"ur Metallforschung,
  D-70506 Stuttgart, Germany}
\author{G.~P.~Mikitik}
\affiliation{Max-Planck-Institut f\"ur Metallforschung,
  D-70506 Stuttgart, Germany}
\affiliation{B.~Verkin Institute for Low Temperature Physics
  \& Engineering, National Ukrainian Academy of Sciences,
  Kharkov 61103, Ukraine}

\date{\today}

\begin{abstract}
We give a theoretical description of the general critical states
in which the critical currents in type-II superconductors are not
perpendicular to the local magnetic induction. Such states
frequently occur in real situations, e.g.,  when the sample
shape is not sufficiently symmetric or the direction of the
external magnetic field changes in some complex way.
Our study is restricted to the states in which flux-line cutting
does not occur. The properties of such general critical states
can essentially differ from the well-known properties of the
usual Bean critical states. To illustrate our approach, we
analyze several examples. In particular, we consider the critical
states in a slab placed in a uniform perpendicular magnetic
field and to which two components of the in-plane magnetic
field are then applied successively. We also analyze the critical
states in a long thin strip placed in a perpendicular magnetic
field which then is tilted towards the axis of the strip.
\end{abstract}


\maketitle

\section{Introduction}  

The concept of the critical state introduced by Charles Bean
\cite{1} is widely used to describe various physical phenomena in
the vortex phase of type-II superconductors, see, e.g.,
Refs.~\onlinecite{2,3} and citations therein. According to Bean,
in the critical state of type-II superconductors with flux-line
pinning, the driving force of the currents flowing in this state
is balanced by the pinning force acting on the vortices. The
critical state is characterized by the component of the current
density flowing {\it perpendicular} to the flux lines,
$j_{c\perp}$, since only this component generates a driving force.
It is assumed in the critical-state theory that this $j_{c\perp}$
is known, i.e., it is a given function of the magnetic induction
${\bf B}$, $j_{c\perp}=j_{c\perp}({\bf B})$, and the problem of
this theory is to find the appropriate distribution of the
magnetic fields and currents in the critical state. Below, for
simplicity, we shall assume that the magnetic fields ${\bf H}$ in
the superconductor considerably exceed the lower critical field
$H_{c1}$, and so we put ${\bf B}=\mu_0 {\bf H}$ throughout the
paper. Beside this, we deal only with bulk superconducting
samples, assuming that all their dimensions noticeably exceed the
London penetration depth, and we consider the critical state
macroscopically, averaging vortex structures and the appropriate
microscopic currents over a scale exceeding the intervortex
spacing.

Hereafter we shall call the critical states ``Bean critical
states'' if the current density ${\bf j}$ is perpendicular to the
local magnetic field ${\bf H}$ at every point of a superconductor,
${\bf j}={\bf j}_{\perp}$, and thus $j=j_{\perp}=j_{c\perp}$. This
definition imposes limitations on the direction of the currents in
the critical state, but it does not imply constancy of
$j_{c\perp}$, e.g., $j_{c\perp}(H)$ can be as in the Kim model.
\cite{kim} The Bean critical states can be found from the static
Maxwell equations,
\begin{eqnarray}\label{1} 
{\rm rot}{\bf H}={\bf j},\ \ \  {\rm div}{\bf H}=0,
\end{eqnarray}
and the conditions on the current density
\begin{eqnarray}\label{2} 
 {\rm div}{\bf j}=0, \ \ \ \ \   \\
 \label{3}  j_{\perp}=j_{c\perp},\ \ \
j_{\parallel}=0,
\end{eqnarray}
where $j_{\parallel}$ is the component of the current density
along the local magnetic field ${\bf H}$. Such states usually
occur when the shape of the superconductor is sufficiently
symmetric and the external magnetic field ${\bf H}_a$ is applied
along a symmetry axis, so that the direction of the currents is
dictated by the symmetry of the problem. Most of the known
solutions of the critical state problem describe just these Bean
states. For example, this is the well-known solution for an
infinite slab in an external magnetic field parallel to its
surface, \cite{1} and also the solution for an infinitely long
cylinder with arbitrary cross-section in a magnetic field parallel
to its axis since the currents flow perpendicular to this axis.
\cite{2} Bean critical states also occur in infinitely long and
thin strips \cite{eh1,eh2,z} and in thin disks \cite{kuz} in a
perpendicular magnetic field even if $j_{c\perp}$ depends on
$|{\bf B}|\equiv B$ or on the angle between ${\bf B}$ and the
normal to the sample plane. \cite{disk,exa,trans} If the applied
magnetic field is tilted to the plane of an infinitely long strip
\cite{zh,obl,obl1,asym} or slab \cite{diana} but remains
perpendicular to the sample axis, the critical currents flow along
this axis, and a Bean critical state occurs. Further examples of
the Bean critical states in samples of a complex shape can be
found in Ref.~\onlinecite{Camp,Pr1,Pr2,Vand1,Vand2}. A
characteristic feature of all these Bean critical states is that
the perturbation of the current distribution caused by a change of
the applied field propagates into the sample as a sharp front at
which the direction of the currents changes abruptly.

In real samples of nonsymmetric shape, or when the applied
magnetic field changes not only in amplitude but also in its
direction, adjacent flux lines may be slightly rotated relative to
each other in the critical state. This rotation generates a
component of the current {\it along} the magnetic field,
\cite{Clem} ${\bf j}_{\parallel}$. The rotation of flux-lines can
lead to their mutual cutting. \cite{2,Clem} Flux line cutting
occurs when the component of the current density parallel to the
magnetic field, $j_{\parallel}$, exceeds some longitudinal
critical current density $j_{c\parallel}$. In this situation a
vortex \cite{cut1} or a vortex array \cite{cut2} becomes unstable
with respect to a helical distortion, and the growth of this
distortion leads to flux-line cutting. When both $j_{\parallel}$
and $j_{c\perp}$ are equal to their critical values
$j_{c\parallel}$ and $j_{c\perp}$, respectively, the so-called
double critical state \cite{Clem,CP} occurs in the superconductor.
\cite{C} For example, this state appears in some region of a
superconducting sample \cite{CP,CP1} when a rotating magnetic
field of constant magnitude is applied to a superconducting disk
(or slab) in its plane. \cite{rot1,rot2,rot3} The double critical
state can be still described by Eq.~(\ref{1}), (\ref{2}), but with
the following conditions on the current density ${\bf j}={\bf
j}_{\perp}+{\bf j}_{\parallel}$:
\begin{eqnarray}\label{4}
 j_{\perp}=j_{c\perp},\ \ \ j_{\parallel}=j_{c\parallel} .
\end{eqnarray}
The concept of the critical state with flux-line cutting was
further developed in Refs.~\onlinecite{Ya1,Ya2,F1} to explain the
observed suppression of the magnetic moment of a superconducting
slab under the action of an ac magnetic field. \cite{F1,F2,F3}

However, in many real situations a change of the direction of the
external magnetic field or a nonsymmetric shape of the sample does
not lead to flux cutting in the superconductor, i.e.,
$j_{\parallel}$ does not reach $j_{c\parallel}$ in the critical
state. In such situations there is no explicit condition on the
magnitude of $j_{\parallel}$ except that $j_{\parallel}<
j_{c\parallel}$, and the static equations (\ref{1}) and (\ref{2})
with the only restriction $j_{\perp}=j_{c\perp}$  {\it are not
sufficient} to find the distributions of the magnetic field ${\bf
H}({\bf r})$ and current density ${\bf j}({\bf r})$ in the
critical state. This problem for the special case of a slab with
an in-plane magnetic field was solved in
Refs.~\onlinecite{CP,CP1}. The full set of the critical-state
equations for arbitrary shape of the sample and for any
quasistatic evolution of the vector of the applied magnetic field
${\bf H}_a$ was obtained in Ref.~\onlinecite{MB05}, where it was
also shown that in contrast to the common Bean critical states, a
perturbation of the current distribution in such critical states
propagates into the sample smoothly in a diffusive way. We
emphasize that this class of critical states with $j_{\parallel}<
j_{c\parallel}$ corresponds to the general situation, while the
common Bean critical states and the double critical states are
only {\it limiting cases} occurring when $j_{\parallel}=0$ or
$j_{\parallel}= j_{c\parallel}$, respectively.

Such general critical states, which we shall call the T-critical
states (T means transport), \cite{C1} occur even for simple
experimental situations. In particular, they appear in a certain
region of thin rectangular platelets in a perpendicular magnetic
field (in platelets with thickness exceeding the London
penetration depth this is the region which is not penetrated by
the perpendicular component of the magnetic field). \cite{ani}
Critical states of this type also appear at the vortex-shaking in
rectangular platelets \cite{rect} and even in strips if the ac
field is along the axis of the strips. \cite{long} They also occur
in low-frequency ac experiments with a slab when a {\it
circularly} polarized ac field is applied perpendicularly to the
dc magnetic field $H_a$ that is normal to the plane of the slab.
\cite{G1,G2}

As was pointed out in Refs.~\onlinecite{CP,CP1}, one more type of
critical states can exist in superconductors. In these states
$j_{\perp}<j_{c\perp}$ and $j_{\parallel}= j_{c\parallel}$, i.e.,
only flux cutting occurs without any transport of vortices. The
description of such C-critical states (C means cutting) in samples of
arbitrary shape can be obtained by an immediate generalization of
the approach used in Ref.~\onlinecite{CP,CP1} for a
superconducting slab. Below we shall not analyze such states in
detail but only briefly outline this generalization.

In Sec.~II of this paper we develop the approach of
Ref.~\onlinecite{MB05}. In particular, we take into account the
dependence of $j_{c\perp}$ on $j_{\parallel}$ and anisotropy of
flux-line pinning. We also discuss the relationship between the
equations of Ref.~\onlinecite{MB05} and the variational principle
recently proposed. \cite{BL1,BL2,BL3} In Sec.~III we then analyze
three examples of the general T-critical state.

\section{General critical states}   

\subsection{Critical-state equations}  

The critical state is well established in a sample if the
characteristic time of change of the applied magnetic field ${\bf
H}_a$, $j_{c\perp}d/|d{\bf H}_a/dt|$, considerably exceeds the
time of flux flow across the sample, $\mu_0 d^2/\rho_{\rm ff}$,
where $d$ is a characteristic size of the sample and $\rho_{\rm
ff}$ is the flux-flow resistivity. In other words, the concept of
the critical state can be used for a description of the
magnetic-field and current distributions in superconductors if the
generated eddy electric fields are relatively small,
 \begin{equation}\label{5}
 \mu_0 d \, \left| {d{\bf H}_a \over dt}\right| \ll
 \rho_{\rm ff}j_{c\perp}.
 \end{equation}
The ideal critical state thus corresponds to the limit $\rho_{\rm
ff}\to \infty$. Below we imply condition (\ref{5}) to be
fulfilled.

The general T-critical states with $j_{\parallel} <
j_{c\parallel}$ can be described by the following approach:
\cite{MB05} The static equations (\ref{1}) and (\ref{2}) are
supplemented by the quasistatic Maxwell equation
 \begin{equation}\label{6}
{\rm rot}\,{\bf E}=-\mu_0 \dot {\bf H},
 \end{equation}
where $\dot {\bf H}\equiv \partial {\bf H}/ \partial t$, and ${\bf
E}$ is the electric field generated by a change of the applied
field ${\bf H}_a$. For the set of equations (\ref{1}), (\ref{2}),
and (\ref{6}) to be solvable, it has to be supplemented by the
current-voltage law  ${\bf E}({\bf j},{\bf B})$. \cite{LL} This
law is introduced from two well-known physical ideas: $1$) At any
given ${\bf j}$ and ${\bf B}$, the {\it direction} of ${\bf E}$
follows from ${\bf E}=[{\bf B}\times {\bf v}]$, i.e.,
 \begin{equation} \label{7}
{\bf E} \parallel  [{\bf B}\times {\bf v}],
 \end{equation}
where ${\bf v}$ is the vortex velocity caused by the Lorentz force
$[{\bf j}\times {\bf B}]$. Here for simplicity we shall neglect
the so-called Hall angle, \cite{BS} and so the directions of ${\bf
v}$ and the Lorentz force coincide. $2$) The {\it magnitude} of
${\bf E}$ is found from the condition that
 \begin{equation} \label{8a}
|{\bf j}_{\perp}|=j_{c\perp}.
 \end{equation}
In fact, this condition may be interpreted as the following
current--voltage dependence:
 \begin{eqnarray} \label{8}
|{\bf E}|&=&0 \ \ \ \ \ \ \ {\rm at}\ \ j_{\perp}<j_{c\perp},
\nonumber \\
|{\bf E}|&\to& \infty \ \ \ \ \ \ {\rm at}\ \ j_{\perp}>
j_{c\perp},
 \end{eqnarray}
which just corresponds to the ideal critical state.

To proceed with our analysis, let us introduce the following
notations for the magnetic field ${\bf H}({\bf r})$ and the
current density ${\bf j}({\bf r})$ in the critical state: ${\bf
H}({\bf r})=H({\bf r}){\bm \nu}({\bf r})$, ${\bf j}({\bf
r})=j({\bf r}){\bf n}({\bf r})$ where $H$ and $j$ are the absolute
values of the magnetic field and the current density while the
unit vectors ${\bm \nu}$ and ${\bf n}$ define their directions.
Then, the component of the current density perpendicular to the
magnetic field is given by
 \[
{\bf j}_{\perp}= {\bf j}- {\bm \nu}({\bm \nu}{\bf j}) \equiv
j_{c\perp}{\bf n}_{\perp}({\bf r}).
 \]
Here the unit vector ${\bf n}_{\perp}$ defines the direction of
${\bf j}_{\perp}$, ${\bf n}_{\perp}=({\bf n}- {\bm \nu}({\bm
\nu}{\bf n}))/D$; $D= \sqrt{1-({\bf n} {\bm \nu})^2}$ is the
normalizing factor that is equal to the sine of the angle between
${\bf H}$ and ${\bf j}$, and we have taken into account the
condition $|{\bf j}_{\perp}| = j_{c\perp}$. These formulas also
lead to the explicit expression for the magnitude $j$ of the
current density,
 \begin{equation} \label{9a}
 j  = {j_{c\perp}({\bf H})\over D},
 \end{equation}
that is only another form of the condition $|{\bf j}_{\perp}| =
j_{c\perp}$. Let us now formulate condition (\ref{7}). Let at a
moment of time $t$ the external magnetic field ${\bf H}_a(t)$
change infinitesimally by $\delta {\bf H}_a= \dot {\bf H}_a \delta
t$. Under the change of ${\bf H}_a$, the critical currents locally
shift the vortices in the direction  of the Lorentz force $[{\bf
j}\times {\bm \nu}]$; this shift generates an electric field
directed along $[{\bm \nu}\times [{\bf j}\times {\bm \nu}]]={\bf
j}_{\perp}$, i.e., along the vector ${\bf n}_{\perp}$. Thus, we
can represent the electric field ${\bf E}({\bf r})$ in the form:
\begin{equation} \label{10a}
{\bf E}={\bf n}_{\perp}e,
\end{equation}
where the scalar function $e({\bf r})$ is the modulus of the
electric field. Note that the electric field in general is not
parallel to the total current density ${\bf j}({\bf r})$. With
formulas (\ref{9a}) and (\ref{10a}), equations (\ref{1}),
(\ref{2}), and (\ref{6}) are sufficient to describe the T-critical
states in a sample of arbitrary shape. It is important that the
magnetic fields ${\bf H}({\bf r})$ and currents ${\bf j}({\bf r})$
in the critical state at the moment of time $t+\delta t$ depends
only on the field and current distributions in the previous
critical state at the moment $t$ and on the change of the external
field $\delta {\bf H}_a= \dot {\bf H}_a \delta t$, while the
electric field $e$ is proportional to the sweep rate $\dot {\bf
H}_a$ rather than to $\delta {\bf H}_a$, and so it plays an
auxiliary role in solving the critical-state problem. \cite{MB05}

We emphasize that $e$ is now found as a solution of the set of
equations (\ref{1}), (\ref{2}), (\ref{6}), (\ref{9a}), (\ref{10a})
without using the specific current-voltage dependence (\ref{8}).
The explicit equation for the scalar function $e({\bf r})$ has the
form: \cite{MB05}
\begin{eqnarray}\label{11a}
{\bf n}_{\perp}\!\cdot \{ {\rm rot\, rot}({\bf E})- ({\bm
\nu}\cdot {\rm rot}{\bm \nu})\, {\rm rot}({\bf E}) \}= \ \ \ \
\nonumber \\ {\partial j_{c\perp}({\bf H})\over \partial {\bf
H}}\cdot {\rm rot}({\bf E}),
\end{eqnarray}
where ${\bf E}$ is given by Eq.~(\ref{10a}). Continuity of the
magnetic field on the surface of the superconductor, $S$, yields
the boundary condition to Eq.~(\ref{11a}):
  \begin{eqnarray}\label{12a}
  -{\rm rot}\left ({\bf E}({\bf r}_S )\right )
  = \mu_0 \dot {\bf H}_a + \! 
  \int {[{\bf R}\times {\rm rot\, rot}({\bf E}({\bf r'}))]
  \over 4\pi R^3}\, d {\bf r}',~
  \end{eqnarray}
where ${\bf r}_S$ is a point on the surface $S$, ${\bf R}\equiv
{\bf r}_S-{\bf r}'$, $R=|{\bf R}|$, and the integration is carried
out over the volume of the sample. The right hand side of this
boundary condition expresses $\mu_0 \dot{\bf H}$ on the surface of
the superconductor (but reaching from outside) with the use of the
Biot-Savart law. If in the critical state of the superconductor
there are also boundaries at which the direction of the critical
currents changes discontinuously or which separate regions with
$j_{\perp}=j_{c\perp}$ from regions with $j=0$, \cite{C3} the
function $e({\bf r})$ has to vanish at these boundaries to provide
continuity of the electric field $e{\bf n}_{\perp}$ there.

In practical calculations of critical states developing in the
process of changing ${\bf H}_a(t)$ it is convenient to rewrite
Eqs.~(\ref{1}) and (\ref{6}) in the form
 \begin{equation} \label{11}
\mu_0 j [{\bf n}\times \dot {\bf n}] =-[{\bf n}\times {\rm rot\,
rot}({\bf E})],
 \end{equation}
which is a differential equation for the angles defining the
direction of ${\bf j}$, i.e., the unit vector
${\bf n} = {\bf j}/j$. Note that
since the distributions of the magnetic fields and currents in the
critical states of a superconductor are independent of the sweep
rate $\dot {\bf H}_a$, their temporal dependence is only a
parameterization of their dependence on ${\bf H}_a$.

Let us now write explicitly the applicability condition of the
above theory. Since the projection of ${\bf j}$ on the local
direction of ${\bf H}$ is $j_{c\perp}({\bf n}{\bm \nu})/D$, the
condition that flux-line cutting is absent leads to the following
restriction on the angle between the local ${\bf j}$ and ${\bf
H}$:
 \begin{equation}\label{13a}  
{|{\bf n}{\bm \nu}|\over \sqrt{1-({\bf n}{\bm \nu})^2}}<
{j_{c\parallel}\over j_{c\perp}},
 \end{equation}
where $j_{c\parallel}$ is the longitudinal critical current
density.

Finally, we make several remarks on the electric field. It may
turn out that the electric field $e{\bf n}_{\perp}$ obtained with
Eq.~(\ref{11a}) does not satisfy the condition ${\rm div}(e{\bf
n}_{\perp})=0$. To clarify this situation, it is necessary to
remember that a moving vortex generates an electric dipolar
moment, \cite{BS} and hence the moving vortex medium is
characterized by the vector of polarization ${\bf P}$ which is the
macroscopic density of this moment. It follows from the results of
Ref.~\onlinecite{BS} that ${\bf P}=-e{\bf n}_{\perp}$, and a
nonzero ${\rm div}(e{\bf n}_{\perp})$ means that in a type-II
superconductor the electric-charge density $-{\rm div}{\bf P}$
appears which generates a curl-free electric field ${\bf
E}_p=-\nabla \Phi$ described by the scalar potential $\Phi$. This
potential field is a part of the total electric field given by
${\bf E}= e{\bf n}_{\perp}$ inside the sample, and it obeys the
equation ${\rm div}{\bf E}_p={\rm div}(e{\bf n}_{\perp})$, i.e.,
 \begin{equation} \label{14a}   
\Delta \Phi = -{\rm div}(e{\bf n}_{\perp}),
 \end{equation}
where $\Delta \equiv {\rm div} \nabla$. At the surface of the
sample, $S$, the field ${\bf E}_p$ satisfies the same boundary
conditions as in the electrostatics of dielectrics: \cite{LL} The
tangential components of ${\bf E}_p$ and the normal component of
${\bf E}_p+ {\bf P}={\bf E}_p - e{\bf n}_{\perp}$ are continuous
there. Since ${\bf P}=0$ outside the sample, the latter condition
means that
 \begin{equation} \label{14b}
 ({\bf E}_p^+ -{\bf E}_p^-){\bm \tau}=
 -e{\bf n}_{\perp}{\bm \tau}
 \end{equation}
where ${\bf E}_p^+$ and ${\bf E}_p^-$ are the surface potential
fields calculated outside and inside $S$, respectively, and ${\bm
\tau}$ is the normal to $S$ pointing outside. The right hand side of
Eq.~(\ref{14b}) gives the surface-charge density induced by moving
vortices in the sample. Note that the potential part of $e{\bf
n}_{\perp}$ does not influence the magnetic fields and currents in
the critical state since ${\rm rot}{\bf E}_p=0$. Appearance of
this part is caused by condition (\ref{10a}) that dictates the
direction of the electric field in the sample. Although both the
inductive part of the electric field, $e{\bf n}_{\perp}-{\bf
E}_p$, which generates the critical states, and the potential part
${\bf E}_{p}$ can be measured in certain situations, \cite{Jooss}
we shall not analyze electric fields in detail in this paper since
these fields plays only an auxiliary role in the critical state
problem. See also the recent book on electric fields. \cite{book}

Generally speaking, in the process of changing ${\bf H}_a$ a
migration of the induced charges $\rho= {\rm div}(e{\bf
n}_{\perp})$ occurs, which leads to a generation of currents
satisfying ${\rm div}{\bf j}=-(\partial \rho/\partial t)$ and
violating Eq.~(\ref{2}). However, these nonstationary currents are
proportional to the second power of the sweep rate $\dot {\bf
H}_a$ and are negligible under assumption (\ref{5}).

\subsection{Generalizations}  

We now point out some generalizations of the above results which
may be useful in analyzing critical states in real situations.

\subsubsection{$j_{c\perp}$ depends on $j_{\parallel}$}  

The current-voltage law used in Sec.~II A, Eq.~(\ref{8}), means
that flux creep is negligible in our approach.
In this case the critical current
density is found from the condition that the creep activation
barrier $U$ of a vortex bundle is equal to zero. It has been
implied above that $j_{c\perp}$ may depend on ${\bf B}$ but is
completely independent of the magnitude of $j_{\parallel}$. In
other words, the form $U=U(j_{\perp},{\bf B})$ has been assumed
for this $U$. However, the creep activation barrier $U$, generally
speaking, may depend not only on $j_{\perp}$ and ${\bf B}$ but
also on the $j_{\parallel}$ that characterizes flux-line
misalignment in the bundle, i.e., in the general case one has
$U=U(j_{\perp},j_{\parallel},{\bf B})$. Then the critical current
density $j_{c\perp}$ determined from
$U(j_{\perp},j_{\parallel},{\bf B})=0$ takes the form
$j_{c\perp}=j_{c\perp}({\bf B},j_{\parallel})$. One may expect
that this dependence of $j_{c\perp}$ on the longitudinal current
component $j_{\parallel}$ is especially noticeable when
$j_{\parallel}$ is close to its critical value $j_{c\parallel}$,
and hence $j_{c\perp}({\bf B},j_{c\parallel})$ in general differs
from $j_{c\perp}({\bf B},0)$. Similarly, the activation barrier
$U_{\rm cut}$ for flux cutting is a function of both
current-density components and of the magnetic induction, i.e.,
$U_{\rm cut}=U_{\rm cut}(j_{\perp},j_{\parallel},{\bf B})$, and
the condition $U_{\rm cut}=0$ gives $j_{c\parallel}=
j_{c\parallel}({\bf B},j_{\perp})$. In Fig.~1a at a fixed ${\bf
B}$ we schematically show the dependences of $j_{c\perp}$ on
$j_{\parallel}$ and of $j_{c\parallel}$ on $j_{\perp}$ in the
plane ($j_{\perp}$,$j_{\parallel}$). Note that these dependences
cross when the equations $U(j_{\perp},j_{\parallel},{\bf B})=0$
and $U_{\rm cut}(j_{\perp},j_{\parallel},{\bf B})=0$ hold
simultaneously. This occurs at isolated points in the
$j_{\perp}$-$j_{\parallel}$ plane since the barriers $U$ and
$U_{\rm cut}$ characterize different physical processes and are
essentially different functions of the current components. These
points correspond to the double critical states when
$j_{\perp}=j_{c\perp}$ and $j_{\parallel}=j_{c\parallel}$. In
Fig.~1a the top/bottom and right/left sections of the curves
between the four points describe $j_{c\perp}(j_{\parallel})$ in
the general T-critical state and $j_{c\parallel}(j_{\perp})$ in
the C-critical state.

The dependence $j_{c\perp}(j_{\parallel})$ leads to a replacement
of $j_{c\perp}({\bf H})$ by $j_{c\perp}({\bf H},j_{\parallel})$ in
formula (\ref{9a}) that now reads 
 \begin{equation} \label{15a}
 j D  = j_{c\perp}({\bf H},j\sqrt{1-D^2}).
 \end{equation}
The dependence $j_{c\perp}(j_{\parallel})$ also leads to a
modification of Eq.~(\ref{11a}). In the right hand side of this
equation the term $-\mu_0(\partial j_{c\perp}/\partial
j_{\parallel})(\partial j_{c\parallel}/\partial t)$ should be
added that equals
  \begin{equation} \label{15b}
  {\partial j_{c\perp}\over \partial j_{\parallel}}\left
({j_{c\perp}\over H}\ {\bf n}_{\perp}\!\cdot {\rm rot}({\bf
E})+{\bm \nu}\cdot {\rm rot\, rot}({\bf E})\right )
  \end{equation}
with ${\bf E}$ from Eq.~(\ref{10a}). Note that the first term in
this expression has no singularity at $H \to 0$ since the
combination ${\bf n}_{\perp}\!\cdot {\rm rot}(e{\bf n}_{\perp})$
can be also rewritten as $e{\bf n}_{\perp}\!\cdot {\rm rot}{\bf
n}_{\perp}$ and $e=|[{\bf B}\times {\bf v}]| \propto H$.

In Refs.~\onlinecite{PR1,PR2} a phenomenological model of the
general critical state was considered that described sufficiently
well a number of experimental data on the magnetization of a slab
and of a disk in magnetic fields parallel to their planes. In
fact, in this model a certain type of the dependence of
$j_{c\perp}$ on $j_{\parallel}$ (and of $j_{c\parallel}$ on
$j_{\perp}$) was introduced. Even though the directions of the
electric field in this model do not satisfy the physical
requirement (\ref{10a}), the sufficiently good description of the
data seems to indicate the importance of this dependence in real
situations.

 \begin{figure}  
 \includegraphics[scale=.47]{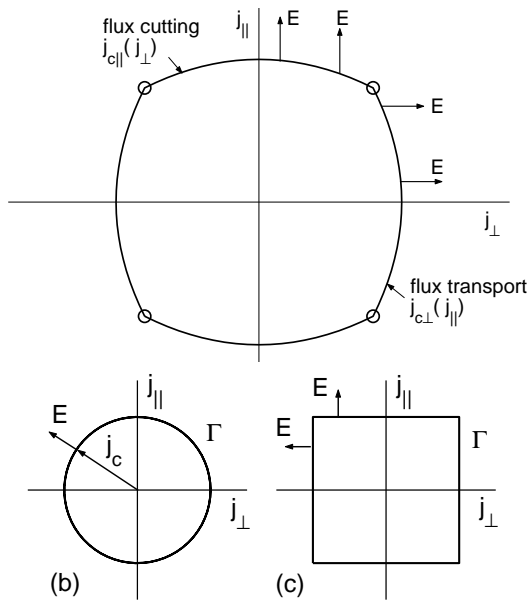}
\caption{\label{fig1} (a): Dependences of  $j_{c\perp}$ on
$j_{\parallel}$ and of $j_{c\parallel}$ on $j_{\perp}$ (solid
lines) shown schematically in the $j_{\perp}$-$j_{\parallel}$
plane at a fixed ${\bf B}$, see Sec.~II B1. The crossing points of
the lines correspond to the double critical states when
$j_{\perp}=j_{c\perp}$ and $j_{\parallel}=j_{c\parallel}$. Shown
are also the directions of the electric field for the appropriate
critical states. (b) and (c): The boundaries $\Gamma$ (solid
lines) of the regions $\Delta$ introduced by Bad\'ia and L\'opez
\cite{BL1,BL2,BL3} in the $j_{\perp}$-$j_{\parallel}$ plane, see
Sec.~II C. Here the regions $\Delta$ are (b) a circle and (c) a
rectangle. Shown are also the directions of the electric
field according to the approach of Bad\'ia and L\'opez.
 } \end{figure}   

\subsubsection{Anisotropy of flux-line pinning}  

In deriving Eq.~(\ref{10a}) we have assumed that when ${\bf H}_a$
changes, vortices shift in the direction  of the local Lorentz
force $[{\bf j}\times {\bf B}]$. However, in the case of
anisotropic pinning this assumption may fail. Nevertheless, even
in this case the direction of the shift can be expressed via the
directions of ${\bf j}$ and ${\bm \nu}\equiv {\bf H}/H$, see
Appendix A in Ref.~\onlinecite{ani}. Now the unit vector ${\bf u}$
along the electric field, ${\bf E}={\bf u}e$, is
 \begin{equation}\label{13}
 {\bf u}={\bf n}_{\perp}\cos \delta + [{\bm \nu}\times {\bf
 n}_{\perp}]\sin\delta ,
 \end{equation}
where the angle $\delta$ describes the change of the direction of
the electric field due to anisotropic pinning. If in the plain
perpendicular to the local ${\bf H}$ the critical current density
$j_{c\perp}$ depends on its direction ${\bf n}_{\perp}$, the angle
$\delta$ is found from \cite{ani}
 \begin{equation}\label{14}
\tan\delta =-{\partial [\ln j_{c\perp}({\bf B}, \phi)] \over
\partial \phi} \,,
 \end{equation}
where $\phi$ is the angle defining the direction of ${\bf
n}_{\perp}=(\cos\phi,\sin\phi)$ in the plane perpendicular to
${\bf H}$. When $j_{c\perp}$ is isotropic in this plane, we obtain
$\delta=0$, and thus ${\bf u}$ coincides with ${\bf n}_{\perp}$.

Equations (\ref{13}) and (\ref{14}) give a relation between ${\bf
n}_{\perp}$ and ${\bf u}$. When $\delta \neq 0$, i.e., when the
vector ${\bf u}$ differs from ${\bf n}_{\perp}$, the only change
in the critical state equations is that $e{\bf n}_{\perp}$ in
Eq.~(\ref{10a}) is replaced by $e{\bf u}$, and $j_{c\perp}({\bf
H})$ in Eq.~(\ref{9a}) is now $j_{c\perp}({\bf H},{\bf
n}_{\perp})$.

\subsubsection{C-critical states}  

As it was mentioned in the Introduction, in the case of an
infinite slab the critical states with flux-line cutting but
without flux-line transport were considered in
Refs.~\onlinecite{CP,CP1}. For samples of arbitrary shape such
C-critical states can be described by Eqs.~(\ref{1}), (\ref{2}),
and (\ref{6}), but now the electric field is along the local ${\bf
H}$,  i.e., ${\bf E}={\bm \nu}e$. This condition replaces
Eq.~(\ref{7}) [or (\ref{10a})]. The absolute value $e$ of the
electric field is now determined by the condition
$j_{\parallel}=j_{c\parallel}$ which is equivalent to the
following current-voltage dependence:
 \begin{eqnarray} \label{15}
|{\bf E}|&=&0 \ \ \ \ \ \ \ {\rm at}\ \
j_{\parallel}<j_{c\parallel},
\nonumber \\
|{\bf E}|&\to& \infty \ \ \ \ \ \ {\rm at}\ \ j_{\parallel}>
j_{c\parallel},
 \end{eqnarray}
and leads to the formula
 \begin{equation} \label{18b}
 j  = {j_{c\parallel}({\bf H})\over |{\bf n} {\bm \nu}|}.
 \end{equation}
Equations (\ref{15}), (\ref{18b}) replace Eqs.~(\ref{8}) and
(\ref{9a}), respectively.

\subsection{Variational principle}  

Recently, \cite{BL1,BL2,BL3} a variational principle was put
forward to describe the critical states in superconductors. In
deriving this principle Bad\'ia and L\'opez used Eqs.~(\ref{1}),
(\ref{6}) and the current--voltage law with $|{\bf E}|=0$ when
${\bf j}$ is inside some region $\Delta$ of the ${\bf j}$-space
and $|{\bf E}|\to \infty$ when ${\bf j}$ lies outside this region.
In other words, the critical states correspond to the boundary
$\Gamma$ of the region $\Delta$, see Fig.~1b,c. However, the
physical idea of the {\it direction} of the electric field,
Eq.~(\ref{7}), was not incorporated in their principle. Instead of
this they find the direction of the electric field from some
condition of maximality of their Hamiltonian. This leads them to
the conclusion that the electric fields in the critical states are
directed along the normals to the boundary $\Gamma$ at the
appropriate points, Fig.~1b,c.

Within our approach their boundary $\Gamma$ corresponds to the
contour composed of the dependences $j_{c\perp}(j_{\parallel})$
and $j_{c\parallel}(j_{\perp})$, see Fig.~1a and Sec.~II B1. But
in our general T-critical states with $j_{\parallel}<
j_{c\parallel}$ the electric field is always perpendicular to the
local ${\bf H}$ (i.e., to ${\bf j}_{\parallel}$), and in the
C-critical states with flux-line cutting but without flux-line
transport the electric field is along the local ${\bf H}$. It is
clear that only in the case when $\Delta$ is a rectangle does the
approach of Bad\'ia and L\'opez lead to the correct results for
the electric field, Fig.~1.\cite{C2} However, in general their
approach leads to contradiction with existing physical
concepts.\cite{2,cut1,cut2} In particular, in the so-called
isotropic model, when $\Delta$ is a circle, Fig.~1b, the electric
field ${\bf E}$ is parallel to ${\bf j}$, and hence a nonzero
${\bf E}$ along ${\bf H}$ appears even for an infinitesimally
small longitudinal component of ${\bf j}$, i.e., flux-line cutting
in that model occurs without any threshold $j_{c\parallel}$.
\cite{C2a}

\section{Examples}  

We first consider two examples of the general critical state in an
infinite slab of thickness $d$. Let this slab fill the space
$|x|$, $|y| <\infty $, $|z|\le d/2$, and be in a constant and
uniform external magnetic field $H_{az}$  directed along the $z$
axis, i.e., perpendicularly to the slab plane. The critical
current density $j_{c\perp}$ is assumed to be constant in this
slab. In the first example a constant field $H_{ax}$ ($H_{ax}\ge
J_c/2=dj_{c\perp}/2$) is applied along the $x$ axis, and after
that the magnetic field $H_{ay}$ is switched on in the $y$
direction. This example was considered in our paper, \cite{MB05}
but there $H_{ax}$, $H_{ay}$, and $J_c$ were assumed to be small
as compared with $H_{az}$, i.e., the tilt angle $\theta$ of the
magnetic field to the $z$ axis was always small. Now we do not put
this restriction, and the angle $\theta$ may be sufficiently
large. But we still assume that flux-line cutting does not occur
(see below). This example may be considered as a modification of
the experimental conditions of Refs.~\onlinecite{F1} where the
suppression of the magnetic moment of the slab was investigated at
$H_{az}=0$. In the second example the critical current along the
$y$ axis is applied to a slab, and after that the magnetic field
$H_{ay}$ is switched on in the same direction.

The critical state equations are the same for these two examples.
The difference is only in the boundary conditions. Let us write
these equations. The condition ${\rm div}{\bf j}=\partial
j_z/\partial z=0$ together with $j_z|_{z=\pm d/2}=0$ yields
$j_z=0$, i.e., the currents flow in the $x$-$y$ planes. \cite{C2c}
Then, to describe the critical state, we may use the
parameterization:
\begin{eqnarray}
{\bf j}=j_{c}(\varphi,\theta,\psi) (\cos\varphi(z), \
\sin\varphi(z),0), \nonumber \\
{\bf H}(z)={\bf\hat z}H_{az}+ {\bf h}(z), \label{20a} \\
{\bf h}(z)=(h_x(z), h_y(z),0), \nonumber
\end{eqnarray}
where ${\bf\hat z}$ is the unit vector along the $z$ axis;
$j_c(\varphi,\theta,\psi)$ is the magnitude of the critical
current density when a flux-line element is given by the angles
$\psi$ and $\theta$, $\tan\psi=h_y/h_x$, $\tan\theta =
(h_x^2+h_y^2)^{1/2}/H_{az}$, while the current flows in the
direction defined by the angle $\varphi$; all these angles
generally depend on $z$. A dependence of $j_c$ on the orientation
of the local ${\bf H}$ appears even at a constant $j_{c\perp}$  if
$j_c$ is not perpendicular to this ${\bf H}$, and this dependence
is described by formula (\ref{9a}), where $D$ in terms of the
angles is
 \begin{equation}\label{20b}
D=[1 - \cos^2(\varphi-\psi) \sin^2\theta]^{1/2}.
 \end{equation}

With this parameterization, the equation ${\rm div}{\bf H}=0$ is
satisfied identically, while the Maxwell equation ${\rm rot}{\bf
H}={\bf j}$ reads
\begin{eqnarray} \label{16}
   {d h_x\over dz}={j_{c\perp}\sin\varphi \over D} , \\
   -{d h_y \over dz}={j_{c\perp}\cos\varphi \over D} . \label{17}
\end{eqnarray}
These equations differ from the appropriate equations of
Ref.~\onlinecite{MB05} by the factor $1/D$, which is not unity
now.

In the case under study one has ${\bf n}=(\cos\varphi,
\sin\varphi, 0)$, ${\bm \nu}=(\sin\theta \cos\psi, \sin\theta
\sin\psi, \cos\theta)$. Then, a direct calculation gives the
following expressions for the vector ${\bf n}_{\perp}$ defining
the direction of the current component perpendicular to ${\bf H}$:
\begin{eqnarray}
 n_{\perp x}&=&{\cos\varphi-\sin^2\theta \cos\psi \cos(\varphi-\psi)
 \over D},  \nonumber \\
n_{\perp y}&=&{\sin\varphi-\sin^2\theta \sin\psi
\cos(\varphi-\psi) \over D},
 \label{17c} \\
n_{\perp z}&=&-{\sin\theta \cos\theta \cos(\varphi-\psi) \over D},
 \nonumber
\end{eqnarray}
and equation (\ref{11a}) for the electric field $e$ takes the
form:
 \begin{eqnarray}\label{18}
\!\!\!\!\!n_{\perp x}(en_{\perp x})''\!\!\!&+&\!\!n_{\perp
y}(en_{\perp y})''
 \ \ \ \ \ \ \ \ \ \ \ \nonumber\\
 &-&\!\!\psi'\sin^2\theta (n_{\perp x}'n_{\perp y}-n_{\perp
x}n_{\perp y}')e =0 ,
 \end{eqnarray}
where the dash over a symbol means $\partial/\partial z$. For the
angle $\varphi$ we obtain from Eq.~(\ref{11}):
\begin{eqnarray}\label{19}
 \mu_0 {j_{c\perp}\over D}{\partial \varphi \over \partial t}=
 (e n_{\perp y})'' \cos\varphi - (e n_{\perp x})'' \sin\varphi \,.
\end{eqnarray}
At small $\theta$ when $D\approx 1$ and ${\bf n}_{\perp}\approx
{\bf n}$, equations (\ref{18}) and (\ref{19}) reduce to the form
that was used in Ref.~\onlinecite{MB05}:
\begin{eqnarray}\label{18a}
 e''-(\varphi')^2 e&=&0 , \\
 \mu_0 j_{c\perp}{\partial \varphi \over \partial t}&=&
 2e'\varphi' +e\varphi'' \,. \label{19a}
\end{eqnarray}

Equations (\ref{16})-(\ref{19}) provide the complete description
of {\it any} general T-critical state in an infinite slab with
$j_{c\perp}({\bf H})=$ constant in the absence of flux-line
cutting. For different critical-state problems only the boundary
conditions should be appropriately chosen. Note that the usual
Bean critical states in the slab correspond to discontinuous
solutions $\varphi(z)$ of these equations.

In the case of the slab condition (\ref{13a}) of absence of
flux-line cutting leads to the following restriction on the angles
$\theta$, $\psi$, and $\varphi$:
 \begin{equation}\label{20}
{\sin\theta |\cos(\varphi -\psi)|\over D}< {j_{c\parallel}\over
j_{c\perp}}.
 \end{equation}
This condition is fulfilled at any $\varphi$ and $\psi$, i.e., at
any direction of ${\bf j}$ and ${\bf h}$, if the $z$ component of
the magnetic field, $H_{az}$, is not too small,
 \begin{equation}\label{21}
\tan \theta = {\sqrt{h_x^2+h_y^2}\over H_{az}} <
{j_{c\parallel}\over j_{c\perp}}\,.
 \end{equation}
We imply this condition to be fulfilled below.

\subsection{First example: $H_{ax}$ and $H_{ay}$}  

In the first example that we consider, a constant field $H_{ax}$
($H_{ax}\ge J_c/2=dj_{c\perp}/2$) is applied along the $x$ axis,
and after that the magnetic field $H_{ay}$ is switched on in the
$y$ direction. Then, the boundary conditions to Eqs.~(\ref{16}) -
(\ref{19}) at $z=d/2$ are
\begin{eqnarray}\label{22}
  h_x=H_{ax},\ \ \  h_y=H_{ay}(t) , \\
  (en_{\perp x})'=-\mu_0 {dH_{ay}(t)\over dt},\ \ \
  (en_{\perp y})'=0,  \label{23}
\end{eqnarray}
or equivalently, conditions (\ref{23}), which follow from formula
(\ref{12a}), can be rewritten in the form:
\begin{eqnarray}\label{24}
  e'(n_{\perp x}'n_{\perp y}-n_{\perp x}n_{\perp y}')=
  \mu_0 {dH_{ay}(t)\over dt} n_{\perp y}', \nonumber \\
  e(n_{\perp x}'n_{\perp y}-n_{\perp x}n_{\perp y}')=
  -\mu_0 {dH_{ay}(t)\over dt}n_{\perp y}.
\end{eqnarray}
Taking into account the symmetry of the problem, 
$e(-z)=e(z)$, $\varphi(-z)=\varphi(z)-\pi$, ${\bf h}(-z)={\bf
h}(z)$, it is sufficient to solve equations (\ref{16})-(\ref{19})
in the region $0 \le z \le d/2$. At $z=0$, where the direction of
the currents changes discontinuously, one has the additional
condition for $e$,
\begin{equation}\label{25}
  e(0)=0.
\end{equation}
Since after switching on $H_{ax}$, the critical currents flow in
the $y$ direction, we have the following initial condition for
Eq.~(\ref{19}):
\begin{equation}\label{26}
  \varphi (z,t=0)=\pi/2 ,
\end{equation}
where the moment $t=0$ corresponds to the beginning of switching
on $H_{ay}$. As to the initial magnetic-field profiles, equations
(\ref{16}), (\ref{17}), and (\ref{26}) give $h_y(z,t=0)=H_{ay}=0$
and $h_x(z,t=0) =H_{ax} -0.5J_c +j_{c\perp}z$ where $J_c \equiv
j_{c\perp}d$.

 \begin{figure}  
 \includegraphics[scale=.47]{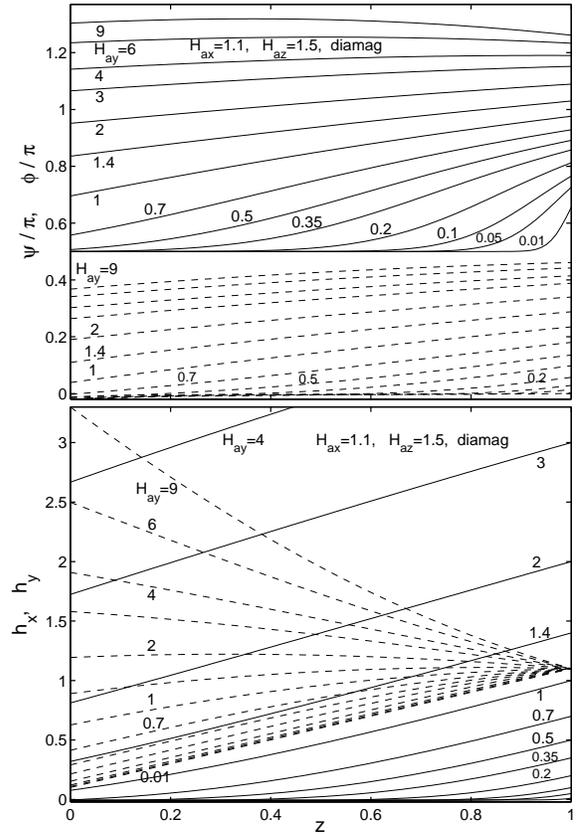}
\caption{\label{fig2} Top: Profiles of the angle of the currents
$\varphi(z)$ (solid lines), and of the field angle $\psi(z)$,
$\tan\psi= h_y(z)/h_x(z)$ (dashed lines) in the critical states of
the slab at $H_{az}=1.5$, $H_{ax}=1.1$ and $H_{ay} = 0.01$,
$0.05$, $0.1$, $0.2$, $0.35$, $0.5$, $0.7$, $1$, $1.4$, $2$, $3$,
$4$, $6$, $9$. Bottom: The magnetic field components $h_x(z)$
(dashed lines) and $h_y(z)$ (solid lines) in the same critical
states. We start from the {\it diamagnetic} initial critical state
with $h_x(z)=H_{ax}-1+z$, $h_y(z)=H_{ay}=0$, and $\varphi(z) =
\pi/2$. Here $z$ is in units of $d/2$, and the magnetic fields in
units of $j_{c\perp} d/2=J_c/2$.
 } \end{figure}   

In the limiting case $H_{az} \gg H_{ax}$, $H_{ay}$, $J_c$, the
solution of equations (\ref{16})-(\ref{19}) with conditions
(\ref{22}) - (\ref{26}) was investigated in
Ref.~\onlinecite{MB05}. Since in this case ${\bf n}_{\perp}\approx
{\bf n}$, one finds that the electric field $e{\bf n}_{\perp}$ is
along the current density $j{\bf n}$, and in fact, we arrive at a
situation which can be formally described by the so-called
isotropic model of Bad\'ia and L\'opez. \cite{BL2} As was
explained in Sec. II C, this model in general does not lead to the
correct direction of the electric field. In particular, it fails
in the following situation discussed by Bad\'ia and L\'opez:
\cite{BL2} A slab with $H_{az}=0$, $H_{ax}=\,$constant, and
oscillating $H_{ay}$. But in the case $H_{az} \gg H_{ax}$,
$H_{ay}$, $J_c$, which in reality was not considered in
Ref.~\onlinecite{BL2}, the isotropic model gives the correct
results, and the numerical data of Ref.~\onlinecite{BL2} agree
\cite{C2b} with those of Ref.~\onlinecite{MB05} and can be used to
describe this limiting situation.

In the case $H_{az} \sim H_{ax}$, $H_{ay}$, $J_c$, the solution of
equations (\ref{16})-(\ref{19}) with conditions (\ref{22}) -
(\ref{26}) is shown in Fig.~2. We present $\varphi(z)$, $\psi(z)$,
$h_x(z)$, $h_y(z)$ in the sequence of the critical states
developed in the process of increasing $H_{ay}$. We do not show
the electric field $e$ that is proportional to the sweep rate
$\dot H_{ay}$ and  plays an auxiliary role. As was noticed
previously, \cite{MB05} in stark contrast to the Bean critical
states, in which any change of the current direction occurs inside
a narrow front, in the general T-critical state the change of the
angle $\varphi(z)$ with increasing $H_{ay}$ has {\it diffusive
character\/}. But there is a difference between the data of Fig.~2
and the results \cite{MB05} obtained in the case $H_{az}\gg
H_{ax}$, $H_{ay}$, $J_c$ when the currents in the critical state
are almost perpendicular to the local magnetic fields. In the
latter case at $H_{ay}> J_c$ the field profile $h_x(z)$ becomes
practically constant and coincides with $H_{ax}$, while the angle
$\varphi$ tends to $\pi$. On the other hand, we see from Fig.~2
that for $H_{az} \sim J_c$ the angle $\varphi$ lies in the
interval $\pi < \varphi <3\pi/2$ at $H_{ay}> J_c$. In other words,
the $y$ component of ${\bf j}(z)$ has the opposite direction as
compared with the initial state. This leads to the fact that at
$H_{ay}>J_c$ the field $h_x$ {\it increases} towards the central
plane of the slab, $z=0$ (but $h(z)= \sqrt{h_x^2+h_y^2}$ still
decreases towards this plane), and the initial diamagnetic state
with the magnetic moment $M_x=-j_{c\perp}d^2/4$ (per unit area)
turns into a paramagnetic state with positive $M_x$.

 \begin{figure}  
 \includegraphics[scale=.47]{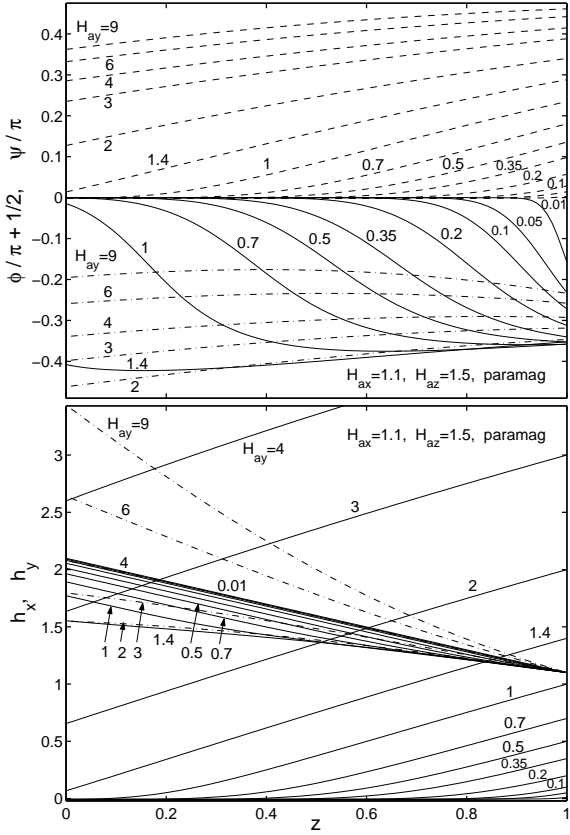}
\caption{\label{fig3} The same as Fig.~2, but at $H_{ay} = 0$ we
start from the {\it paramagnetic} initial critical state with
$h_x(z) = H_{ax}+1-z$, $h_y(z)=0$, and $\varphi(z)= -\pi/2$. For
clarity the profiles $\varphi(z)$ and $h_x(z)$ at $H_{ay} \ge 2$
are shown by dash-dotted lines (and for $H_{ay} \le 1.4$ by solid
lines).
 } \end{figure}   

In Fig.~3 we show the same sequence of the critical states but in
the case of the paramagnetic initial state. This initial state is
obtained if one first increases the field $H_{ax}$ essentially
above the field of full flux penetration and then decreases it to
a prescribed value. Now the initial condition to Eq.~(\ref{19}) is
\begin{equation}\label{26a}
  \varphi (z,t=0)=-\pi/2 ,
\end{equation}
and the magnetic fields at $t=0$ are given by
$h_y(z,t=0)=H_{ay}=0$, $h_x(z,t=0)=H_{ax}+0.5J_c-j_{c\perp}z$. It
is seen from Fig.~3 that although at $H_{ay} < H_{az}$ a decay of
the initial paramagnetic profile $h_x(z)$ occurs, with a
further increase of $H_{ay}$ new paramagnetic states are developed
that are close to the appropriate states of Fig.~2.

In Fig.~4 we compare the $H_{ay}$-dependences of the magnetic
moment ($M_x$, $M_y$) per unit area of the slab, \cite{c4}
 \begin{equation}\label{26b}
       {\bf M}=\int_{-d/2}^{d/2}z[{\bf\hat z}\times {\bf j}]dz,
 \end{equation}
obtained using the two sequences of the profiles ${\bf
j}(z,H_{ay})$ developed from the diamagnetic and paramagnetic
initial states with the same $H_{ax}$. It is seen that in both
cases $|M_x|$ and $|M_y|$, and even more $M = \sqrt{M_x^2
+M_y^2}$, can exceed the ``saturation value'' $j_{c\perp}d^2/4$
used as unit in Fig.~4. This is possible since the current density
$j$ exceeds $j_{c\perp}$ when it does not flow at a right angle to
the vortices. This excess of $j$ leads to that $M_x(H_{ay})$ does
not saturate at large $H_{ay}$ but continues to increase nearly
linearly, with slightly negative curvature. The other component,
$M_y(H_{ay})$, at large $H_{ay}$ practically saturates to a value
slightly lower than $-j_{c\perp}d^2/4$. Of course, one should keep
in mind that in reality the region of large $H_{ay}$ where these
results for ${\bf M}$ are applicable is limited by condition
(\ref{21}). Note also that in agreement with Figs.~2 and 3 the
magnetic moment $M_x(H_{ay})$ is always positive at sufficiently
large $H_{ay}$, and the diamagnetic and paramagnetic initial
states lead to practically the same $M_x$ at such $H_{ay}$.

 \begin{figure}  
 \includegraphics[scale=.47]{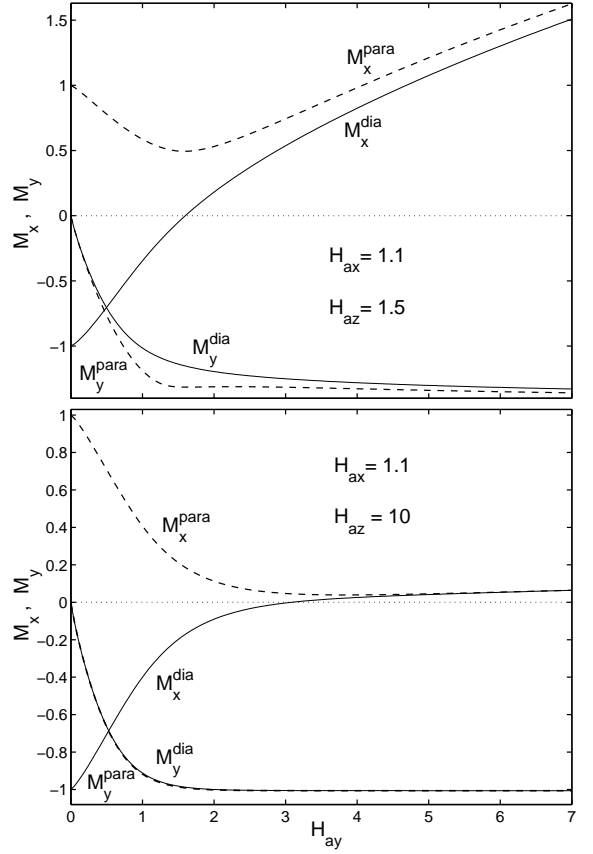}
\caption{\label{fig4} The magnetic moment $(M_x, M_y)$ of the slab
per unit area defined by Eq.~(\ref{26b}) as a function of
$H_{ay}$. Shown are the diamagnetic (solid lines, see Fig.~2) and
paramagnetic (dashed lines, see Fig.~3) cases for $H_{ax}=1.1$ and
$H_{az}=1.5$ (top) and $H_{az}=10$ (bottom). Units are
$j_{c\perp}d/2$ for $H$ and $j_{c \perp} d^2/4$ for $M$.
 } \end{figure}   

\begin{figure}  
  \includegraphics[scale=.47]{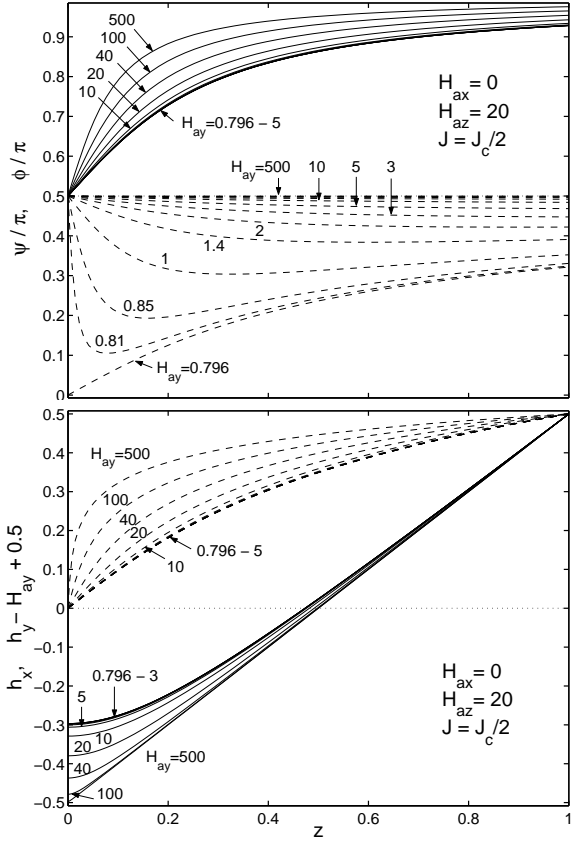}
\caption{\label{fig5} Profiles of the angles $\varphi(z)$ (solid
lines) and $\psi(z)$ (dashed lines) (top),  and of the magnetic
field components $h_x(z)$ (dashed lines) and $h_y(z)$ (solid
lines) (bottom) for a slab with applied sheet current $J = J_c/2$
and $H_{ax}=0$. The profiles are calculated from Eqs.~(\ref{30})
and (\ref{31}) for $H_{ay}=0.796$, $0.81$, $0.85$, $1$, $1.4$,
$2$, $3$, $5$, $10$, $20$, $40$, $100$, $500$ and large
$H_{az}=20$ (all the magnetic fields are in unit of $J_c/2$).
$H_{ay}^0=0.796$ is the field of full flux penetration at $J =
J_c/2$.
  } \end{figure}   

As it is known, field-cooled type-II superconducting samples
frequently exhibit a positive magnetic moment; see, e.g., paper
\cite{R1} and references therein. Different explanations of this
paramagnetic effect were put forward. In particular, this effect
may be associated with the compression of trapped magnetic flux in
the sample. \cite{LarKosh} The data of Figs.~2-4 show that in
principle, the paramagnetic effect may be also due to the
field-cooling caused generation of critical states in which the
circulating currents are not perpendicular to the local magnetic
fields.

The general T-critical states considered here can be realized in
experiments similar to the experiments of Park {\it et
al}.\cite{PK} and Fisher {\it et al}., \cite{F1} except that now
the field $H_{az}$ perpendicular to the plane of the sample is not
equal to zero. Such investigations would enable one to compare the
theoretical results for the general T-critical states with the
appropriate experimental data avoiding complications due to
flux-line cutting. To prepare the initial state which was
described above, e.g., in a superconducting strip of length $2L$
and width $2w$ considerably exceeding its thickness $d$, $2L>2w\gg
d$, one may apply first the field $H_{az}$ perpendicular to the
plane of the strip, and then an {\it oscillating} in-plane
magnetic field $H_{ax}$ across the width of the strip. This
``shaking'' leads to a homogeneous distribution of the
perpendicular field $H_{az}$ over the sample. \cite{tran} After
this shaking process one keeps $H_{ax}=$ constant and applies the
field $H_{ay}$ along the axis of the strip.

\subsection{Second example: $J_y$ and $H_{ay}$}  

We now consider the second example of the general T-critical state
in the slab. It is assumed that the slab is in a uniform
magnetic field $H_{az}$ along the $z$ axis, the current $J$ (per unit
length along $x$) flows in the $y$ direction, and at $t=0$ the
field $H_{ay}$ is switched on. In this case the boundary
conditions at $z=d/2$  are
\begin{eqnarray}\label{27}
  h_x={J\over 2},\ \ \  h_y=H_{ay}(t) , \\
  (en_{\perp x})'=-\mu_0 {dH_{ay}(t)\over dt},\ \ \
  (en_{\perp y})'=0.  \label{28}
\end{eqnarray}
The symmetry of the problem is now described by the relationships:
$e(-z)=e(z)$, $\varphi(-z)=\pi-\varphi(z)$, $h_x(-z)=-h_x(z)$,
$h_y(-z)=h_y(z)$, and at $z=0$ the direction of the critical
currents changes continuously. Thus, instead of condition
(\ref{25}) we have at $z=0$,
\begin{equation}\label{29}
  h_x(0)=0\,, \ \ \ \varphi(0)=\pi/2.
\end{equation}
As in the first example, we shall consider the critical states
only in the interval $0\le z \le d/2$.

If $J$ is less than $J_c=j_{c\perp}d$, in the initial state the
current flows only at $1-(J/J_c)\le 2z/d \le 1$. After switching
on $H_{ay}$ the current distribution develops over the whole
thickness $d$ when $H_{ay}$ reaches a penetration field $H_{ay}^0
< J_c/2$, and we shall analyze the critical states only after this
penetration of the current has occurred, i.e., at $H_{ay}\ge
H_{ay}^0$.

Below we consider only the case $H_{az}\gg J_c$. In this case in
the leading order in the small parameter $J_c/H_{az}$ we find the
following analytic solution of Eqs.~(\ref{16}) - (\ref{19}) with
boundary conditions (\ref{27})-(\ref{29}), see also Fig.~5:
\begin{eqnarray}\label{30}
 e&=&{\mu_0\dot H_{ay}a\, D\over \sin\varphi \cos\theta},\ \ \ \
 \cot \varphi=- {z\cos\theta\over a},  \nonumber \\
  h_x&=&{j_{c\perp}a\over \cos\theta} \ln\left ({z\over a}+
  \sqrt{1+{z^2\over a^2}}  \right), \\
   h_y&=&H_{ay}- j_{c\perp}a \left (\sqrt{1+{d^2\over 4a^2}}-
 \sqrt{1+{z^2\over a^2}}\right ), \nonumber
 \end{eqnarray}
where $D^2=1 - \cos^2(\varphi-\psi) \sin^2\theta \approx 1-
\sin^2\varphi \sin^2\theta $ (since either $\psi \approx \pi/2$ or
$\sin^2\theta \ll 1$), $\cos^2 \theta \approx H_{az}^2/(H_{az}^2 +
H_{ay}^2)$, and the length $a$ is determined by the sheet current
$J$ and $\cos\theta$,
\begin{equation}\label{31}
\!\!\!\!\!{J\over J_c}\cos\theta\!=\!{2a\over d}\ln\!\! \left
(\!{d\over
  2a}+\sqrt{1+{d^2\over 4a^2}}\,\right )\!\!=\!{2a\over d}\,
  {\rm arcsinh} {d\over 2a}\,.
\end{equation}
We shall denote the solution of Eq.~(\ref{31}) as $2a/d=g(J
\cos\theta /J_c)$. The function $g(u)$ defined by $u = g\, {\rm
arcsinh} (1/g)$ increases monotonically with its argument $u$,
Fig.~6. Hence with increasing $H_{ay}$, i.e., with decreasing
$\cos\theta $, the length $a$ decreases. When $J$ is close to
$J_c$ and $\cos\theta \approx 1$, the length $a$ tends to
$\infty$, while for $J \ll J_c$ one has $2a/d \sim J
\cos\theta/J_c \ll 1$. A very good approximation valid at all $u$
is, \cite{long}
 \begin{eqnarray} \label{32}
 \nonumber
 g(u)&\approx & u \Big/ \ln
   {(1+g_0^2)^{1/2}+1 \over g_0} \,,  \nonumber \\
   g_0(u)& = &{u + u^2 \over
      [ 24 (1-u) ]^{1/2}} \,.
 \end{eqnarray}
The field of full flux penetration can be estimated from
$h_y(0)=0$,
\begin{equation}\label{33}
H_{ay}^0= j_{c\perp}a \left ( \sqrt{1+{d^2\over 4a^2}}-1 \right ),
\end{equation}
where $a$ is determined by Eq.~(\ref{31}). When $H_{ay}\ll
H_{az}$, one has $\cos\theta \approx 1$, and the length $a$ is
almost independent of $H_{ay}$. Thus, at such $H_{ay}$ the
profiles $\varphi(z)$, $h_x(z)$, and $h_y(z)-h_{ay}$ practically
do not change with increasing $H_{ay}$. Fig.~5 shows that this
property of the profiles, in fact, holds in the region $H_{ay} <
H_{az}/2$ when $J/J_c=0.5$. However, if $J$ is close to $J_c$, the
length $a$ sharply depends on $\cos\theta$, Fig.~6, and the width
of this region shrinks.

 \begin{figure}[b]  
 \includegraphics[scale=.43]{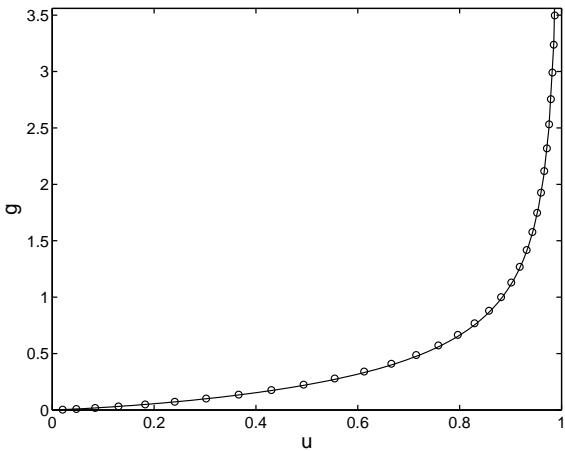}
\caption{\label{fig6} The function $g(u)$ defined by $u = g\, {\rm
arcsinh} (1/g)$ (circles). Also shown is approximation (\ref{32})
(solid line).
 } \end{figure}   

 \begin{figure}[t]  
 \includegraphics[scale=.43]{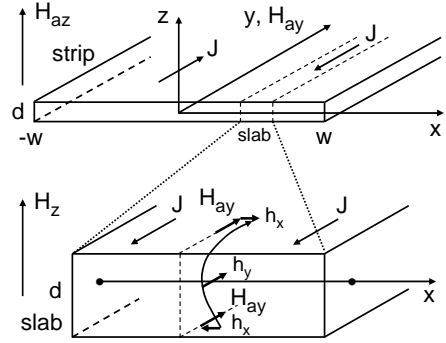}
\caption{\label{fig7} Schematic picture of the strip. Shown is
also a ``slab'' cut-out near the point $x$, see Sec. III C.
 } \end{figure}   

Solution (\ref{30}) can be obtained as follows: We put
 \begin{equation}\label{34}
(en_{\perp x})''=0, \ \ \ (en_{\perp y})''=0,
 \end{equation}
since it may be verified that the term proportional to
$\psi'\sin^2\theta$ in Eq.~(\ref{18}) and the left hand side of
Eq.~(\ref{19}) are small in the parameter $J_c/H_{az}$ and hence
may be omitted in the first approximation. Equations (\ref{34})
mean that $en_{\perp x}$ and $en_{\perp y}$ are linear functions
of $z$, and thus each of them generally depends on two constants.
However, taking into account boundary conditions (\ref{28}) and
the symmetry of the problem, one finds that the functions
$en_{\perp x}$ and $en_{\perp y}$ are expressed by only one
constant that coincides with $en_{\perp y}$. If we denote this
constant as $a\cos\theta$ and use
 \begin{equation}\label{35}
 n_{\perp x}\approx {\cos\varphi \over D}, \ \ \ \ n_{\perp y}
\approx {\sin\varphi \cos^2\theta \over D},
 \end{equation}
we arrive at formulas (\ref{30}). The profiles $h_x(z)$ and
$h_y(z)$ follow from Eqs.~(\ref{16}) and (\ref{17}), and the
constant $a$ can be found from the condition
 \begin{equation}\label{36}
  J=J_y=\int_{-d/2}^{d/2}{dz\, j_{c\perp}\sin\varphi \over D},
 \end{equation}
which is just Eq.~(\ref{31}).

It is also instructive to write the electric-field components $E_x
\equiv en_{\perp x}$ and $E_y \equiv en_{\perp y}$ explicitly.
Using Eqs.~(\ref{30}) and (\ref{35}), we find
\begin{eqnarray} \label{37}
 E_x\!\!&=&-\mu_0\dot H_{ay}z    \\
 E_y\!\!&=&\mu_0\dot H_{ay}a \cos\theta = {\mu_0\dot H_{ay}d\over 2}
 \cos\theta g(J \cos\theta /J_c).\ \ \label{38}
 \end{eqnarray}
The field $E_x$ results from the tilt of a vortex line along the
$y$ direction when $H_{ay}$ is applied to the slab. Note that
$\int_{-d/2}^{d/2}E_x(z)dz=0$ since the upper ($z>0$) and lower
($z<0$) parts of the vortex move in opposite directions when
the tilt occurs. On the other hand, $E_x$ is independent of $z$.
This component of the electric field is due to a drift of the
vortex as a whole in the $x$ direction when $H_{ay}$ is applied to
the sample. \cite{long}

The above formulas for the slab with a current enable one to
reproduce a number of results for the vortex-shaking effect that
were derived from geometrical considerations.\cite{long,rect} In
particular, the expression for $\varphi(z)$ in Eqs.~(\ref{30}),
formula (\ref{31}), and Eq.~(\ref{38}), in fact, coincide with
Eqs.~(4), (6) and (28) from Ref.~\onlinecite{long} in which the
so-called longitudinal vortex-shaking effect in a thin strip was
considered. To obtain the formulas for the vortex-shaking effect
in a rectangular platelet, \cite{rect} one should consider the
slab with $H_{az}\gg H_{ay}$, $J_c$ and with the total current $J$
flowing at an arbitrary angle to the $y$ axis, i.e., when ${\bf
J}=(J_x,J_y)$. The appropriate solution of the critical state
equations is still obtained from Eqs.~(\ref{34}), but now there is
no more symmetry restriction on the $z$ dependences of all the
functions, and $en_{\perp x}$ and $en_{\perp y}$ are expressed via
two constants. Similarly to Eq.~(\ref{36}), these constants can be
expressed via $J_x$ and $J_y$, and the solution thus obtained
reproduces the appropriate results of Ref.~\onlinecite{rect}.

\subsection{Third example: strip}  

We now consider the third example of the general T-critical state.
Let a thin strip fill the space $|x|\le w$, $|y| <\infty $,
$|z|\le d/2$ ($d\ll w$), and be in a constant and uniform external
magnetic field $H_{az}$ directed along the $z$ axis, i.e.,
perpendicularly to the strip plane. The critical current density
$j_{c\perp}$ is still assumed to be constant, and let $H_{az}$
considerably exceed $J_c=j_{c\perp}d$ so that at the initial
moment of time, $t=0$, the strip is in the fully penetrated Bean
critical state. In other words, the magnetic-field profile
$H_z(x)$ in the strip is described by the well-known function,
\cite{eh1,eh2,z} and one has $J_y(x)=J_c$ for $-w\le x <0$ and
$J_y(x)=-J_c$ for $w\ge x >0$, where the sheet current $J_y$ is
the current density integrated over the thickness $d$. At $t>0$
the magnetic field $H_{ay}$ is switched on in the $y$ direction,
and hence the applied field is tilted {\it towards the axis} of
the strip. Note that the critical states in isotropic and
anisotropic strips placed in inclined magnetic fields were studied
in Refs.~\onlinecite{zh,obl,obl1,asym,diana}. However, in all
these papers the external magnetic field was tilted {\it
perpendicularly to the axis} of the strip, the currents in the
critical states were always perpendicular to local magnetic
fields, and thus, the usual Bean critical states occurred in the
strips. In the considered case the general T-critical states
develop in the strip, and these states differ from the states of
the second example in that the magnetic field $H_z$ and the
currents $J$ are not uniform in the $x$ direction any more.

 \begin{figure}  
 \includegraphics[scale=.47]{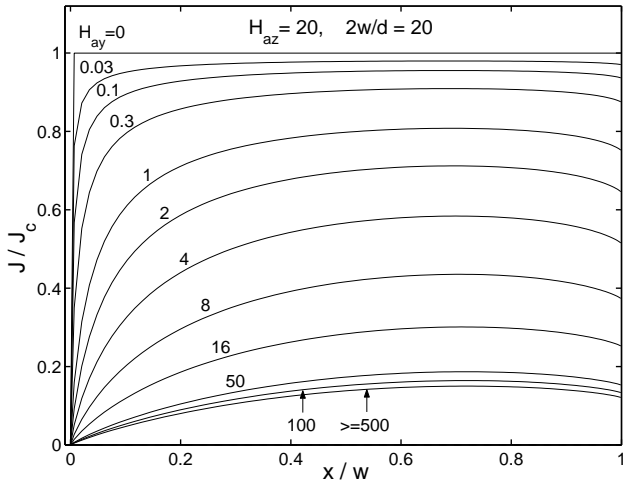}
\caption{\label{fig8}
  The sheet current $J(x)\equiv |J_y(x)|$ in the strip to which first a large
  perpendicular magnetic field $H_{az}=20$ is applied and then an
  increasing longitudinal field $H_{ay}$. The aspect ratio of
  the strip is $2w/d = 20$. The magnetic fields are in units of
  $J_c=j_{c\perp}d$.
 } \end{figure}   

 \begin{figure}  
 \includegraphics[scale=.47]{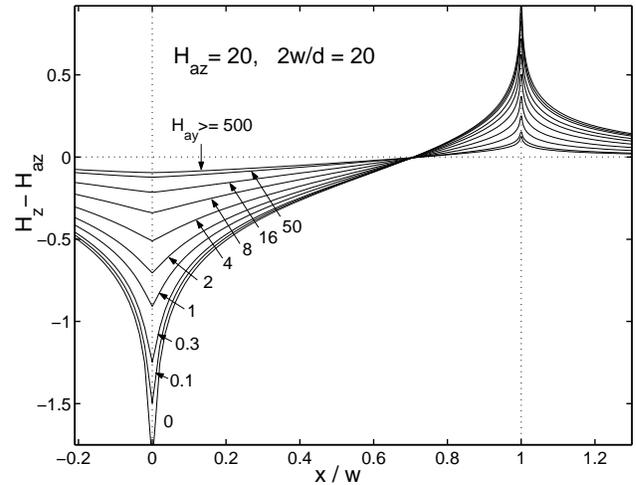}
\caption{\label{fig9} The perpendicular magnetic field $H_z(x)$
caused by  the sheet current of Fig.~\ref{fig8}. The magnetic
fields are in units of $J_c$.
 } \end{figure}   

Strictly speaking, the description of the magnetic-field tilt
towards the axis of the strip reduces to solving a two-dimensional
general T-critical state problem. However, the smallness of the
parameter $d/w$ enables us to simplify this problem by application
of the approach of Ref.~\onlinecite{ani}. Within this approach, we
split the problem into two simpler problems: A one-dimensional
problem across the thickness of the sample, and a problem for the
infinitely thin strip. Namely, we first interpret a small section
of the strip around an arbitrary point $x$ (see Fig.~7) as an
``infinite'' slab of thickness $d$ placed in a perpendicular dc
magnetic field $H_z(x)$ and in a parallel field $H_{ay}$ and
carrying a sheet current $J_y(x)$. This is just the problem that
has been solved in Sec.~III B. We then use the resulting electric
field $E_y$ obtained for the slab, Eq.~(\ref{38}), as the local
electric field $E_y(x)$ for an infinitely thin strip, to calculate
the temporal evolution of the sheet current $J_y(x)$ and of the
magnetic field $H_z(x)$ in this strip by the method of
Ref.~\onlinecite{ehstr1,ehstr2}. The resulting equation for
$J_y(x,t)$ can be written in the form: \cite{tran}
 \begin{equation} \label{39} 
  {\partial J_y(x,t) \over \partial t}={2 \over \pi \mu_0}
  \int_{-w}^w \!
{du \over  u\!-\!x} \left({w^2\!-\!u^2 \over
w^2\!-\!x^2}\right)^{\!\!1/2}
  \! {\partial E_y(J_y) \over \partial u} \,,
 \end{equation}
where $E_y(J_y)$ is given by Eq.~(\ref{38}). On determining
$J_y(x,t)$, the magnetic-field profiles are found from the
Biot-Savart law. Since $E_y(J_y)\propto \dot H_{ay}$, we see again
that the temporal dependence of the current and magnetic-field
profiles is only a parameterization of their dependence on
$H_{ay}$, Sec.~II A. It also follows from Eqs.~(\ref{38}) and
(\ref{39}) that these profiles depend on the parameters $H_{ay}$,
$H_{az}$, $d$, $w$ via the following combinations: $J_y =
J_y(x/w,H_{ay}/H_{az},P)$, $H_z = H_z(x/w,H_{ay}/H_{az},P)$ where
we have introduced the notation $P\equiv (d/2w)H_{az}/J_c$. Note
that the considered critical state problem is similar to the
problem of the longitudinal vortex-shaking effect in a thin strip.
\cite{long} The difference between the problems is that the
magnetic field $H_{ay}$ now increases {\it monotonically} rather
than oscillates about $H_{ay} =0$, and here we present results up
to large values of $H_{ay}$ even as compared with $H_{az}$.

In Figs.~8 and 9 we show the profiles $J(x,H_{ay})\equiv
|J_y(x,H_{ay})|$ and $H_z(x,H_{ay})$ that develop in the strip
during increase of the longitudinal field component $H_{ay}$,
i.e., when the applied field is tilted away from the $z$ axis
towards the strip axis $y$. The profiles $J(x,H_{ay})$ take the
shape similar to the shape of the profiles in the longitudinal
vortex-shaking effect \cite{long}, and their magnitude decreases
with increasing $H_{ay}$. However, in contrast to the shaking
effect, this magnitude does not decrease down to zero but tends to
a finite limit that depends on the only parameter
$P=(d/2w)(H_{az}/J_c)$. Thus, at $H_{ay}\gg H_{az}$ the current
profiles $J(x,H_{ay})$ and the magnetic-field profiles
$H_z(x,H_{ay})$ reach nonzero limiting distributions. The
existence of such limiting $J(x)$ and $H_z(x)$ can be understood
from the following considerations: At small $\cos\theta$, if one
neglects logarithmic corrections, the electric field $E_y$,
Eq.~(\ref{38}), is proportional to $\dot
H_{ay}(d/2w)(J/J_c)\cos^2\theta$, and equation (\ref{39}) has a
solution with separable variables: $J_y(x,H_{ay})=J_c f(x/w)
F(H_{ay})$ where $f(x/w)$ and $F(H_{ay})$ are some functions and
$\cos^2\theta =H_{az}^2/(H_{az}^2+H_{ay}^2)$. Inserting this form
of $J_y(x,H_{ay})$ into Eq.~(\ref{39}), we find that
 \begin{equation}  \label{40}
\ln F(H_{ay}) \propto -P \arctan {H_{ay}\over H_{az}}+ {\rm
const}.,
 \end{equation}
i.e., at $H_{ay} \to \infty$ the function $F(H_{ay})$ does not
tend to zero. In other words, with increasing $H_{ay}$ the decay
rate of $J$ decreases so quickly that $J$ does not reach zero even
in the limit $H_{ay} \to \infty$.

 \begin{figure}  
 \includegraphics[scale=.47]{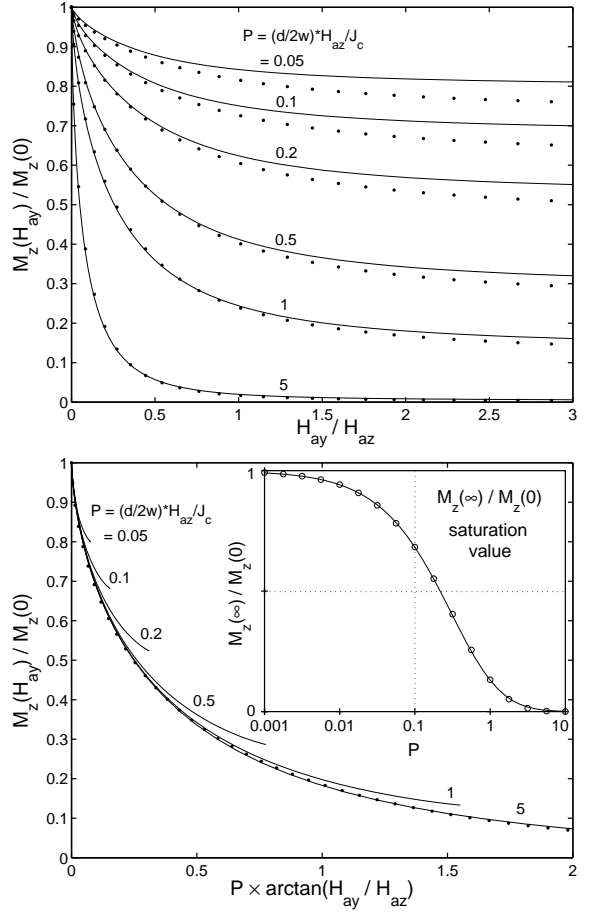}
\caption{\label{fig10} The perpendicular component $M_z$ of
the magnetic moment of the strip shown in Fig.~7, plotted versus
$H_{ay}/H_{az}$ (top) and versus $P\cdot \arctan(H_{ay}/H_{az})$
(bottom) for different values of the parameter $P\equiv
(d/2w)H_{az}/J_c=0.05$, $0.1$, $0.2$, $0.5$, $1$, $5$. The dots
show approximation (\ref{42}). The inset shows the saturation
values $s =M_z(\infty) / M_z(0)$ calculated numerically (circles)
and their fit by Eq.~(\ref{43}) (solid line).
  } \end{figure}   

 \begin{figure}  
 \includegraphics[scale=.47]{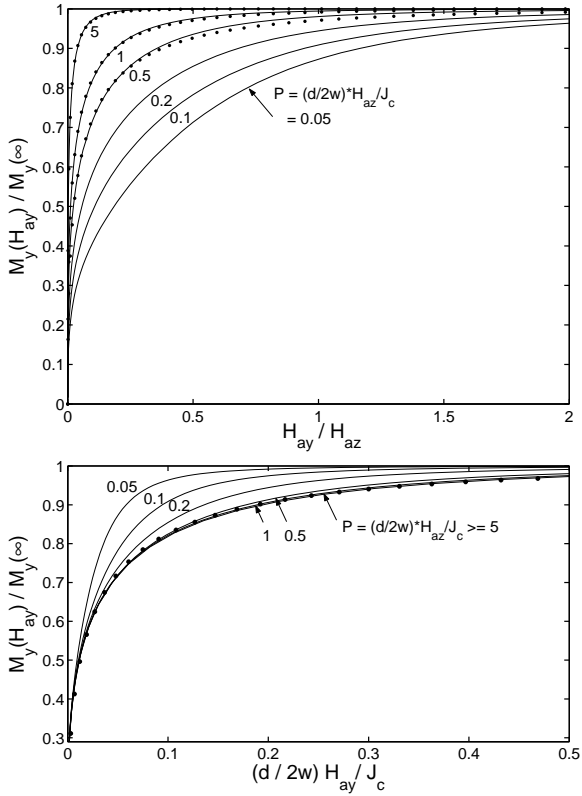}
\caption{\label{fig11} The longitudinal component $M_y$ of
the magnetic moment of the strip shown in Fig.~7, plotted versus
$H_{ay}/H_{az}$ (top) and versus $P\cdot H_{ay}/H_{az}=
(d/2w)H_{ay}/J_c$ (bottom) for the same $P$ as in Fig.~10.
Here $M_y(\infty)=M_y^0=-j_{c\perp}d^2w/2$. Note that the curves
for different $P \ge 0.5$ practically coincide if the
magnetic moment is plotted versus $P\cdot H_{ay}/H_{az}$. The
dots indicate the fit (\ref{47}) for this universal curve.
 } \end{figure}   

In Fig.~10 we show the magnetic moment per unit length of the
strip, $M_z$, in the general T-critical states developed during
increase of $H_{ay}$,
     \begin{equation}  \label{41}  
  M_z(H_{ay}) = -2 \int_{0}^w \! x |J_y(x,H_{ay})|\, dx \,.
     \end{equation}
Here $J_y(x,H_{ay})$ is the solution of Eq.~(\ref{39}), see
Fig.~8. At the initial moment of time, when the usual Bean
critical state occurs, we have $M_z(0)=-J_cw^2$. The application
of $H_{ay}$ leads to the relaxation of $M_z$ towards a saturation
value $M_z(\infty)$. The normalized moment $M_z(H_{ay})/M_z(0)$
depends only on $H_{ay}/H_{az}$ and the parameter $P$. All the
curves of Fig.~10 may be well fitted by stretched exponentials of
the form $s + (1-s) \exp[-p (H_{ay}/H_{az})^q ]$, with some $s$
and $p$ depending on $P$ and with $q$ close to $2/3$. \cite{C5}
However, an expression useful for all $P$ is suggested by Fig.~10
(bottom), namely,
   \begin{equation}\label{42}  
     {M_z(H_{ay}) \over M_z(0)} \approx \exp\{ -1.67 \,
     [ P \arctan(H_{ay}/H_{az}) ]^{0.65} \} \,.
   \end{equation}
This expression, depicted in Fig.~10 (top and bottom) as dots,
gives excellent fits to the numerical results for not too small $P
\ge 0.5$, but even for smaller $P$ it is qualitatively correct and
only slightly underestimates the exact $M_z$ at large
$H_{ay}/H_{az}$.

The saturation values $s =M_z(\infty) / M_z(0)$ are determined by
the above-mentioned limiting current profiles and depend on the
only parameter $P=(d/2w)H_{az}/J_c$. These values obtained
numerically are plotted in the inset of Fig.~10 as circles, while
the solid line in this inset is the following analytic
approximation,
   \begin{equation}\label{43}
 s(P) \approx
   \big( 0.5 -0.5 \tanh \{ 0.41[ \ln(P)-0.5 ] \} \big)^4 \,.
   \end{equation}

When the magnetic field $H_{ay}$ is switched on, not only does the
$z$ component $M_z$ of the magnetic moment change but there appears
also a magnetic moment $M_y$ along the axis of the strip. This
moment (per unit length of the strip) is defined by the
expression:
     \begin{equation}  \label{44}  
  M_y(H_{ay}) = \int_{-w}^w \!dx \int_{-d/2}^{d/2} \!dz j_x z ,
     \end{equation}
where the $x$ component of the current density, $j_x=j_{c\perp}
\cos\varphi /D$, can be found using solution (\ref{30}). With
Eq.~(\ref{17}), formula (\ref{44}) can be rewritten in the form:
    \begin{equation}  \label{45}  
M_y(H_{ay})=-\int_{-w}^w \!dx \int_{-d/2}^{d/2} \!dz
    [H_{ay}-h_y(x,z)].
    \end{equation}
In other words, $M_y$ is the ``expelled'' flux in the $y$
direction. Inserting Eqs.~(\ref{30}) into this formula, we
obtain
\begin{equation}\label{46}  
  {M_y(H_{ay}) \over M_y^0}= \int_0^{w} {dx\over w}\,
  [\sqrt{1+g^2(u)}-ug(u)],
\end{equation}
where $M_y^0=-j_{c\perp}d^2w/2$ is the magnetic moment in the
fully penetrated Bean critical state which occurs if the field
$H_{ay}$ alone is applied to the strip, $u=J\cos\theta/J_c$, and
$J=J(x,H_{ay})$ is the current profile obtained from
Eq.~(\ref{39}), see Fig.~8.

Figure 11 (top) shows the normalized magnetic moment $M_y(H_{ay})
/ M_{ay}( \infty )$ plotted versus $H_{ay}/H_{az}$ for the same
values of the parameter $P$ as in Fig.~10. The saturation value
$M_{ay}( \infty )$ always coincides with $M_y^0=
-j_{c\perp}d^2w/2$. Beside this, we find numerically the following
interesting result: If $P$ is not too small, $P \ge 0.5$, the
normalized magnetic moment plotted versus $P \cdot
H_{ay}/H_{az}=(d/2w) H_{ay}/J_c$ is well described by the unique
curve, Fig.~11 (bottom),
   \begin{equation}\label{47}
    { M_y(H_{ay}) \over M_y(\infty)} \approx 1 - \exp\{ -4.8\,
     [ (d/2w) (H_{ay}/J_c) ]^{0.44} \} \,.
   \end{equation}
At smaller $P$ values fits of the form (62) are still possible,
but with different fitting parameters.

The results of this section describe the relaxation of
$J(x,H_{ay})$ and $H_z(x,H_{ay})$ to the limiting profiles and of
$M_z(H_{ay})$ and $M_y(H_{ay})$ to their saturation values.
According to Figs.~8-11 and Eqs.~(\ref{42}), (\ref{47}), this
relaxation mainly finishes at some $H_{ay}$ proportional to
min$(H_{az},2wJ_c/d)$ [note that $H_{az}\gg J_c=dj_{c\perp}$ for
Eq.~(\ref{38}) to be valid and for the full flux penetration to
occur in the initial state]. All these results can be verified in
experiments similar, e.g., to those of Refs.~\onlinecite{i1,i2}.
However, we emphasize that in contrast to
Refs.~\onlinecite{i1,i2}, the magnetic-field component $H_{az}$
has to be switched on before $H_{ay}$. This guarantees absence of
flux-line cutting for not too large $H_{ay}$, see Eq.~(\ref{21}).
If similarly to experiments \cite{i1,i2} the in-plane magnetic
field is switched on before $H_{az}$, completely different
critical states will develop.

\section{Conclusions}  

In this paper we point out how to calculate the general
T-critical (cutting-free)
states in an arbitrarily-shaped type-II superconductor when the
applied magnetic field ${\bf H}_a$ slowly changes in its magnitude
and direction. In accordance with the definition of the general
T-critical state, it is assumed here that the external magnetic
field changes in such a manner that flux cutting does not occur
in the sample. Our approach enables one to take into account the
anisotropy of flux-line pinning and the dependence of the critical
current density perpendicular to a local magnetic field,
$j_{c\perp}$, on the longitudinal component of the current density
$j_{\parallel}$. We also show that the variational principle
recently proposed \cite{BL1,BL2,BL3} cannot give the correct
description of the general T-critical states for many situations.

We analyze three examples of the general T-critical states, at
least two of which may be investigated experimentally. In
particular, we study a seemingly simple problem that has not been
solved as yet, viz., we consider the critical states in a slab
placed in a uniform perpendicular magnetic field $H_{az}$ and
then two components of the in-plane magnetic field, $H_{ax}$ and
$H_{ay}$, are applied successively, Sec.~III A. We obtain that one
of the in-plane components of the magnetic moment, $M_x$, becomes
positive with increasing $H_{ay}$ for any sign of $M_x$ in the
initial state (i.e., at $H_{ay}=0$). This paramagnetic effect is
due to the fact that the currents in the critical states are not
perpendicular to the local magnetic fields. This effect is
especially evident when $H_{az}$ is of the order of the
self-fields of the slab.

In the other example, we analyze the general T-critical states in
a long thin strip placed in a perpendicular magnetic field
$H_{az}$ which then tilts towards the axis of the strip $y$,
Sec.~III C. When $H_{ay}$, the axial component of the applied
magnetic field, increases, the magnetic-field and current
profiles across the width of the strip tend to limiting profiles,
and the components of the magnetic moment, $M_z$ and $M_y$, reach
saturation values. The limiting profiles and the saturation value
$M_z(\infty)$ for $M_z(H_{ay})$ are determined by the only
parameter $P=(d/2w)H_{az}/J_c$ where $d$ and $2w$ are the
thickness and the width of the strip, respectively, and
$J_c=dj_{c\perp}$. If $P$ is not too large, $P < 5$, the limiting
profiles and $M_z(\infty)$ noticeably differ from zero, while at
$P\ge 5$ they become very small and practically vanish. The
saturation value for $M_y$ is always equal to
$M_y^0=-j_{c\perp}d^2w/2$.

\acknowledgments

  This work was supported by the German Israeli Research Grant
Agreement (GIF) No G-901-232.7/2005.


\begin{thebibliography}{1}
\bibitem{C} Note that in contrast to some papers we use the term
  ``double critical state'' only for the states in which both the
  transverse and longitudinal critical current densities are equal
  to their critical values.
\bibitem{C1} This definition is a generalization of the definition
   used in Ref.~\onlinecite{CP} for the case of a slab.
\bibitem{C3} When ${\bf H}_a$ changes, these boundaries may shift
   in the superconductor. The new positions of the boundaries are
   found from the  continuity of ${\bf H}$ in the sample.
\bibitem{C2} Their approach is also correct for the Bean critical
   states with a fixed direction of the critical currents. 
\bibitem{C2b} Strictly speaking, we used $j_{c\perp}\!\!=\,$const.
in our paper, \cite{MB05} while in Ref.~\onlinecite{BL2}
$j_{c\perp}$ was described by the Kim model. \cite{kim} But for
the same $H$-dependence of $j_{c\perp}$ the appropriate equations
exactly coincide.
\bibitem{C5} Interestingly, the temporal relaxation of the magnetic
  moment in the longitudinal vortex-shaking effect is also
  approximated by a stretched exponential with $q=0.64$. \cite{long}
\end{thebibliography}
\end{document}